\documentclass[reqno,11pt]{article}
\usepackage[]{graphicx}
\usepackage{amsfonts,amssymb,amsmath,amscd,cite}

\def\be{\begin{equation}}
\def\beq{\begin{equation}}
\def\ee{\end{equation}}
\def\eeq{\end{equation}}
\def\ba{\begin{eqnarray}}
\def\ea{\end{eqnarray}}
\def\eqn#1{\begin{equation}\begin{split}#1\end{split}\end{equation}}

\def\P{\partial}

\def\a{\alpha}

\def\text#1{{\hbox{#1}}}
\def\1{{-1}}
\def\2{\frac{1}{2}}
\def\<{\langle}
\def\>{\rangle}
\def\b{\beta}
\def\e{{\epsilon}}
\def\g{\gamma}
\def\l{\lambda}

\def\k{\kappa}

\def\vev#1{\left\<#1\right\>}

\def\IP{\relax{\rm I\kern-.18em P}}

\begin{document}

\vskip.8cm
\vskip.7cm
\begin{Large}
\centerline{{\bf Open-Closed String Correspondence} }  
\centerline{{\bf in Open String Field Theory}}  
\end{Large}
\vskip.2cm
\centerline{by}
\vskip.2cm
\centerline{{\bf 
Marco Baumgartl\footnote{E-mail: \sf baumgartl@itp.phys.ethz.ch}}
}
\centerline{and}
\centerline{{\bf
Ivo Sachs\footnote{E-mail: \sf ivo@theorie.physik.uni-muenchen.de}}
}
\vskip 5mm
\centerline{$^1$\it Institut f\"ur Theoretische Physik, ETH Z\"urich}
\centerline{\it CH-8093 Zurich, Switzerland}
\centerline{$^2$\it Arnold-Sommerfeld Center for Theoretical Physics (ASC)}
\centerline{\it Ludwig-Maximilians-Universit\"at M\"unchen, D-80333 M\"unchen}
\vskip.7cm
\vskip1.2cm
\centerline{{  \bf Abstract} }
\par\noindent
\vskip.8cm
We address the problem of describing different closed string backgrounds in  background independent open string field theory: A shift in the closed string background corresponds to a collective excitation of open strings. As an illustration we apply the formalism to the case where the closed string background is a group manifold.



\vfill\eject

\section{Introduction}
A long standing issue in string theory is whether open strings or closed strings are more fundamental. The simplicity and perturbative completeness of  closed string theory alone might suggest that closed strings are fundamental with open strings arising as excitations of D-branes which are in turn viewed as defects in closed string theory. The problem with this approach is that it is not clear how to write down mutual consistency conditions for D-branes in closed string theory: there are many types of D-branes which are perfectly consistent by themselves but mutually inconsistent with each other. Another problem is that the construction of closed string field theory is a very challenging enterprise \cite{Zwiebach:1997fe} while the construction (bosonic) open string is much more manageable. For instance in Witten's cubic open string field theory \cite{Witten:1985cc} such a definition is encoded in an associative algebra equipped with a BRST operator $Q$ and an invariant scalar product. On the other hand, if one takes open string field theory as a starting point one has to explain how the closed sting poles arising in loop diagrams are contained in the Hilbert space of open string field theory. This is one of the major open problem in string field theory (see \cite{Sen:2004nf} for s recent review and references and \cite{Lazaroiu:2000rk, Kapustin:2004df} for related approaches within topological string theory). 

In this talk we discuss a related problem, that is how to implement closed string backgrounds in background independent open string field theory (BSFT) \cite{Witten:1992qy, Witten:1992cr, Shatashvili:1993kk, Shatashvili:1993ps}. Basically we will argue that certain closed string deformations, corresponding to massless closed string excitations can be implemented in background independent open string field theory in terms of some coherent open string excitation involving an infinite number of open string modes. While this does not answer the question of how elementary closed string excitations are contained in the open string Hilbert space, it goes some way in explaining how coherent, or quasi classical states can be represented in background independent open string field theory. It may also help to classify conformal closed string backgrounds in open string field theory.  In the examples discussed below, where the backgrounds are group manifolds, this classification can be understood in terms of the corresponding  group 2-cocycles, but in more general cases it is not clear what the correct description is. The fact that quasi classical rather than elementary closed string excitations are accessible in BSFT is related to the definition of BSFT in terms of a worldsheet $\sigma$-model on a geometric background. What makes BSFT suitable for studying such backgrounds is the fact that the lowest level excitations, in other words,  massless fields can be switched on in this theory without exciting higher level or massive fields in the open string sector. 

On the other hand BSFT is believed to be related by some yet to be constructed field redefinition to cubic string field theory (OSFT). The latter is conceptually better understood but not background independent. Nevertheless there have been attempts to describe closed string excitations in terms of the deformation theory of the underlying associative algebra. Hopefully a clear understanding of these concepts as well as the relation between the two approaches will eventually become available. 

The plan of the talk is as follows: After a short review of BSFT we will review the factorization property of the boundary sigma model partition function which is the key ingredient for the main result, that is the representation of closed string backgrounds in terms of open string excitations. We then illustrate the formalism with several examples and close with a discussion and some speculations.

\section{Closed Strings in BSFT}

In order to investigate the role which closed strings play in open string field theory we start with a classical configuration of string theory. Perturbative string theory in this background can then be described by a conformal field theory (CFT). This is a two-dimensional quantum field theory with $\sigma$-model action. Generally, this $\sigma$ model depends on the  closed string background fields as well as the open strong couplings, although the requirement of conformal invariance imposes strong restrictions on the background. Once we relax these, it is necessary to define a string theory action away from the classical configuration, which allows for off-shell amplitudes. For the open string $\sigma$-model there is a consistent way how to do that.  

In the proposal, that has been put forward in \cite{Baumgartl:2004iy, Baumgartl:2006xb} one starts with the $\sigma$-model action
\eqn{
	I=\int_\Sigma L_p(G_{\mu\nu},B_{\mu\nu}; X^\mu)+\oint_{\partial\Sigma} B(t^I; X^\mu)\ . 
}
The first of the terms is a standard kinetic term used to define the {\it conformal} closed string $\sigma$-model. It depends on the massless background fields $G_{\mu\nu}$ and $B_{\mu\nu}$, which are the graviton background and the Kalb-Ramond field. In addition to the massless closed string modes there are, of course, also massive modes. The inclusion of these in the $\sigma$-model approach is an open problem. We will thus restrict ourselves to massless modes at present\footnote{We will consider bosonic string  field theory in this talk. Extension to the super string is possible although with some subtleties in the definition of a background independent string field theory action for the super string.}. The second term is a boundary integral over a functional $B$ which depends on the complete set of open string couplings $t^I$. The associated vertex operators are given by 
\eqn{
	\frac{\partial B(t^I; X^\mu)}{\partial t^J}\Biggr|_{t^J=0} = V_J(X^\mu)\ .
}

An important assumption is that the string worldsheet $\Sigma$-model without boundary describes a CFT. The inclusion of a boundary with generic  boundary conditions breaks conformal invariance. 

The generating functional of open string correlation functions is given by the path integral expression
\eqn{
	{\cal Z}[t^I] = \int D[X^\mu] e^{-I}\ .
}
It has been shown in [W,S] that a consistent off-shell action can be defined as
\eqn{
	{\cal S}[t^I] = \left(1-\beta^I\partial_I\right) {\cal Z}[t^I]\ .
}
In this expression the $\beta$-functions associated to the couplings $t^I$ appear. At the conformal fixed point $\beta^I=0$, which is just the solution of the equations of motion of ${\cal S}[t^I]$. Therefore critical points of renormalization group flow coincide with classical solutions of string field theory.

In this formulation ${\cal S}[t^I]$ still depends implicitly on the chosen closed string background
\eqn{
	{\cal M} = (G_{\mu\nu}, B_{\mu\nu})\ .
}
The massless fields parametrize the closed string moduli space at lowest order in the level expansion about a flat target space. On the other hand, the cohomology of physical states in the open string sector crucially depends on the point in moduli space where the theory is formulated. Therefore even a slight change in that moduli space can drastically  affect the open string sector. Let
\eqn{
	{\cal \hat M} = (\hat G_{\mu\nu}, \hat B_{\mu\nu})
}
be a second point in that moduli space. With the above introduced spacetime action ${\cal S}$ we can now compare the two different open string field theories. In the present formalism it can be explicitly tracked, how the open string couplings are changed by a shift in the closed string background. Formally, we find the following {\it equality}:
\eqn{
\label{eqnEQ}
	{\cal S}_{\cal M}(t^I) \equiv {\cal S}_{\cal \hat M}(t^I+\Delta t^I; \Delta \tau^J)\ .
}
The right hand side indicates that in the new background the original values of the couplings do not describe a conformal fixed point, but must also be shifted. This to be expected. More surprisingly, one finds that in addition to the couplings that were present in the ${\cal M}$-background, new couplings $\tau^J$ appear. These correspond generically to non-local operators on the boundary of the worldsheet. 

A truly background independent open string field theory cannot distinguish between a shift in the closed string background and a shift in (non-local) boundary couplings. With this interpretation of (\ref{eqnEQ}) in mind it must be conjectured that non-local couplings, representing collective open string excitations, are needed to complement the theory.

\subsection{Factorization of partition functions}

A key ingredient in the derivation of (\ref{eqnEQ}) is the factorization property of path-integrals \cite{Baumgartl:2004iy} which appears in the definition of ${\cal S}$:
\eqn{
\label{eqnBSFT}
	{\cal S}_{\cal M}(t^I) = {\cal Z}_{{\cal M}}^0\hat {\cal S}_{\cal M}(t^I)
}
Here ${\cal Z}_{{\cal M}}^0$ is the D0-instanton partition function.
$\hat{\cal S}_{\cal M}(t^I)$ is described purely in terms 
of the quantum mechanical degrees of freedom of the string map on the boundary.
This property holds provided once the following factorization of the partition function for 2d conformal field theories 
on a manifold with boundary is established 
\eqn{
	\int D[X^\mu]e^{-I_\Sigma} = {\cal Z}^0{\cal Z}^{bndry}[t^I]
}
Here $I_\Sigma$ is a Lagrangian for the world-sheet conformal field theory on a 2d surface 
with boundary, ${\cal Z}^0$ is the D-instanton partition function which is evaluated under a certain fixed D0 boundary condition and ${\cal Z}^{bndry}[t^I]$ is an integral over boundary degrees of freedom but no bulk degrees of freedom involved.  

This factorization can be easily understood in the case of free bosons corresponding to a Euclidean geometry with all closed sting moduli set to zero. A general string map $X^\mu(z,\bar z)$ can always be decomposed uniquely in the following way
\eqn{
	X^\mu(z,\bar z) = X_0^\mu(z,\bar z) + X_b^\mu(f)\ ,
}
where $X_0^\mu(z,\bar z)$ satisfied D0 boundary conditions and $X_b^\mu$ is a functional of the boundary data $f$ of $X^\mu$. When $f$ specifies the value of the string map restricted to the boundary, then $X_b^\mu(f)$ is given by the unambiguous extension of the function $f$ to the interior of the disk, governed by holomorphicity. To be concrete, we set
$X_0|_{\partial\Sigma} =0$ and assume that $f$ is given as a function of the boundary coordinate $\theta$ with the expansion $f(\theta)=\sum f_ne^{in\theta}$. The equations of motions $\Delta X_b^\mu=0$ then dictate the extension $X_b^\mu(z,\bar z) = f_0^\mu + \sum_{n>0}(f_nz^n + f_{-n} \bar z^n)$. When we split $f$ in a positively and negatively moded part $f^\pm$, the Polyakov action for the free boson takes the form
\eqn{
	I = I_0[X_0] + I[f] = \frac{R^2}{4\pi i\alpha'} \int \partial X_0\bar\partial X_0
		+ \frac{R^2}{4\pi i\alpha'} \int \partial f^+\bar\partial f^-\ .
}
Partial integrals have been used to remove all terms that mix the `bulk' and `boundary' fields. Hence it is obvious that the path-integral satisfies
\eqn{
	{\cal Z} = \int D[X_0] e^{-I_0[X_0]} \cdot \int D[f] e^{-I[f]}\ ,
} 
since there is no additional contribution from the measure. By assumption the inclusion of any boundary operators will result in a modification of $I[f] \mapsto I[f] + t^I V_I[f]$ only. It is not necessary to specify the boundary interaction. It is clear that the path-integral factorizes even without conducting the $f$-integration. This is the simplest example of a string theory which satisfies the factorization property. 

The importance of this property lays in the fact that it now makes sense to define the object
\eqn{
	\frac{{\cal Z}}{{\cal Z}^0} = \int D[f] e^{-I[f]}\ .
}
In the definition of the BSFT action (\ref{eqnBSFT}) is it now possible to choose a $t^I$-independent normalization at each point of the closed string moduli space. The factorization property ensures that this normalization can be kept all over the moduli space. It is important to note that as a consequence of the factorization property the resulting field theory of open strings is not an effective theory in the sense that closed string degrees of freedom are integrated out.

\subsection{Constant Kalb-Ramond field}

In the case where the closed string background has flat geometry but is endowed with a Kalb-Ramond field $B_{\mu\nu}$ the factorization can be shown in the same way, as long as the field is constant. Explicitly, the boundary action in the presence of a $B$-field becomes
\eqn{
	I[f] \propto \oint d\theta f^\mu(\theta)\partial_\theta \bar f^\nu(\theta) \delta_{\mu\nu}
		+ \oint d\theta X_b^\mu\partial_\theta X_b^\nu B_{\mu\nu}\ .
}
The local fields at the boundary are given by $X_b[f]=f+\bar f$. Thus the Kalb-Ramond field induces a local terms under the boundary integral. This reflects the fact that in flat non-compact space a constant $B$-field appears in the open string sector like a constant electromagnetic field strength. The metric induced part, ie.\ the first term in the expression, is not local in the same way. It can be re-expressed in terms of the boundary fields in the form
\eqn{
	\oint d\theta X_b^\mu(\theta) (H_{\mu\nu} X_b^\nu)(\theta)\ ,
}
where $H_{\mu\nu}$ denotes the Hilbert transform 
\eqn{
	H_{\mu\nu}(\theta,\theta') = \frac{1}{4\pi i} \sum_n e^{in(\theta-\theta')}|n|\ ,
}
acting on $X_b$ with integration. The existence of this term comes from the fact that the propagator is furnished by the spacetime metric, hence its appearance is generic. In the case of non-constant Kalb-Ramond field more non-local terms appear. A prototype of such theories is given by WZW models, on which we will focus in the next paragraph.

\subsection{Group manifolds}

While the factorization of the path-integral is obvious in the flat boson case, it is generally not clear that it hold also for more complicated target space manifolds. It has been proven in  \cite{Baumgartl:2004iy}  that for strings on group manifolds which can be expressed as WZW-model, the factorization indeed does hold. We will review this briefly.

The string map $X^\mu$ is replaced by a map $g$ into the Lie group $G$ of the WZW model. In spirit similar to the flat boson case, a general such map can be decomposed as product
\eqn{
	g(z,\bar z)=g_0(z,\bar z) k(z,\bar z)\ ,
}
where $g_0=1$ at the boundary of the worldsheet. $k$ is again a map which is defines as continuation of boundary data $f(\theta)$ according to the equations of motion. In general this is a difficult problem. In the WZW model, however, it can be solved by the ansatz
\eqn{
	k(z,\bar z) = h(z)\bar h(\bar z)\ .
}
The boundary action cannot be determined uniquely in this case. This lays in the fact that the topological term in the WZ action has an ambiguous in contribution in the presence of boundaries. This ambiguity is given by an exact 2-form $\omega=d\beta$ which can be pulled back to the boundary of the disk. In the closed string case such a term is not present, but in the open string case it appears as field on the D-brane. We can set this contribution to zero while keeping in mind that the freedom of adding it always persists.
Formally, the boundary contribution is given by
\eqn{
	I_\beta[k]= A[h,\bar h] + \oint \beta\ .
}

The boundary action which appears after dividing out the D-instanton contribution turns out to have the structure of a 2-cocycle. The presence of a boundary on the world sheet give rise to an extension of the loop group $\hat{LG}$ with an exact sequence
\eqn{
	0\to U(1)\to\hat{LG}\to LG \to 1
}
The cocycle relation is
\eqn{
	(\delta A)[a,b,c] = aA[b,c] - A[ab,c] + A[a,bc] - A[a,b]c = 0\ ,
} 
where
\eqn{
	A[a,b] \sim A[a,b] + \delta \oint \omega[a,b]\ .
}
The 2-form $\omega=d\beta$ specifies an open string background. For general values of $\beta$ the action will not be conformal, but it is always possible to choose $\beta$ is way so that the action describes the theory at a conformal fixed point.

Although $A[h,\bar h]$ cannot be written as a (local) boundary integral in general, it still is completely determined by the boundary data $f$ of the string map on the disk. In the next section we will give an example for a WZW model with $S^3$ target.

\subsection{Stability of D-branes in $S^3$}

The BSFT with $S^3\cong SU(2)$ target has been explicitly derived in \cite{Baumgartl:2006xb}. For large values of the compactification radius the theory can approximated by flat space. When no conditions are imposed of the boundary data $f$, the theory describes a three-dimensional spacefilling brane. This is a conformal background in flat space, but at finite radius the $SU(2)$-WZW model is known to posses no spacefilling branes. The only allowed (maximally symmetric) branes are 2-spheres and D-instantons.

In order to be explicit we choose coordinates on $SU(2)$
\eqn{
	h(\theta) = e^{i\lambda f^a(\theta) T_a}\qquad	\bar h(\theta) = e^{-i\lambda \bar f^a(\theta) T_a}\ .
}
The 2-cocycle $A[h,\bar h]$ can be explicitly expressed in the following way
\eqn{
	A[h,\bar h] = \oint_{\partial\Sigma} d\theta \bigl[ f^a\partial_\theta f_a
		+ i\lambda \epsilon_{abc}f^a \bar f^b \partial_\theta f^c + h.c\Bigr] + {\cal O}(\lambda^2)\ .
}
There is also a contribution from the measure in the path-integral that must be taken into account. Perturbatively, it becomes
\eqn{
	D[k] = D[f] D[\bar f] e^{-\lambda^2 \oint f^a\bar f_a +{\cal O}(\lambda^2) }\ .
} 
In fact, this contribution from the measure  can be re-written locally in terms of the boundary maps as $\oint \lambda^2 X_b^2$. This has the interpretation of a quadratic tachyon profile. It is absent at infinite radius $\lambda=0$. But when the theory is formulated at finite, the tachyon appears automatically and cannot be set to zero. Rather the tachyon renders the open string sector unstable. It initiates a renormalization group flow which drives the theory towards a conformal vacuum -- a process known as tachyon condensation.

One can speculate about the end-point of that flow. From comparison with tachyon condensation in flat space it is clear that instantons are possible end-points. More interesting is the question of there are also condensation channels which lead to vacua with 2-dimensional brane. In order to analyze the renormalization group flow we can add suitable open string interactions to the action, such as
\eqn{
	\oint \beta = \rho\oint d\theta ( X_b(\theta)^2-c^2)^2\ .
}
Here $\rho$ and $c$ enter as couplings. Their $\beta$-functions turn out to be
\eqn{
	\beta_\rho &<0 \qquad \text{always}\\
	\beta_{c^2} &= 1-2c^2\l^2 + {\cal O}(\lambda^2)\ ,
} 
where $\lambda$ is the inverse radius of $SU(2)$.

We can make two remarks to these $\beta$-functions: First, the coupling $\rho$ takes the role of the tachyon. It triggers a condensation process which is inevitable, since its $\beta$-function cannot become positive. Second: at the end-point of the condensation ($\rho=\infty$) conformal configurations are obtained by solving the constraints from all other $\beta$-functions. The equations for $c^2$ suggest that
\eqn{
	c^2=\frac{1}{2\lambda^2}
}
describes the end point. 

This fixed point has a geometric interpretation as a 2-dimensional spherical geometry. Since the RG equations have been obtained in perturbation theory, it must be verified that the theory at least posses such an RG flow fixed point. This is possible by imposing the geometrical constraint $\delta(X_b^2-c^2)$ explicitly in the path-integral.
By construction this has the correct geometry. Its conformality can be again confirmed by a computation of the $\beta$-functions. The resulting condition is again $c^2\sim\lambda^{-2}$. Therefore the 2-sphere is indeed a conformal 2-brane. Moreover this is qualitative agreement with exact results \cite{AlekseevMC} on conformal boundary states in the corresponding WZW model.

\subsection{Nappi-Witten background}

Another illustrative example is the WZW model based on a Nappi-Witten background. 
This background describes a homogeneous anisotropic four dimensional space-time 
with signature $(-+++)$  that can be constructed from an ungauged WZW model with central charge $c$ equal to four, so it can be directly substituted for four dimensional 
Minkowski space and regarded as a solution of any more or less realistic string 
theory. It describes a special case of a monochromatic plane wave with more than 
the usual symmetry. 

The action is given by \cite{Nappi:1993ie}
\eqn{
S= S_0+S_1\ ,
}
where
\eqn{
\label{NWA}
S_0 &= \int\Bigl[
		\P_\a a_k\P^\a a^k + 2\P_\a v\P^\a u + b\P_\a u\P^\a u \Bigr]\\
S_1 &= \int\Bigl[
			 \e_{jk}\P^\a u\P_\a a_ja_k + 2i\e_{\b\g}u\P^\b a_1\P^\g a_2
		\Bigr], \
}
where $b$ is a constant. Coordinates on the group are given by
$g=\exp(a_iP^i)\exp(uJ+vT)$. $P^1,P^2,J,T$ form the Lie algebra
\eqn{
	[J,P_i]=\epsilon_{ij}P_j \qquad [P_i,P_j]=\epsilon_{ij}T \qquad [T,J]=[T,P_i]=0\ ,
}
which is a central extension of the two-dimensional Poincare algebra.

Splitting the fields like before with the ansatz $m=gk$ where $k$ classical and $g$ a field with Dirichlet boundary conditions defines a Nappi-Witten action $S(gk)$.
This action is uses an inner product, which is, other than in WZW models over semi-simple groups,
not given by the Killing form, since the latter is degenerate for this algebra. The crucial property needed so that
the Polyakov-Wiegmann formula holds is the equality $\vev{[X,Y],Z} = \vev{X, [Y,Z]}$, where $X,Y$ and $Z$ are elements of the algebra. It can be checked that this property is satisfied in our case. There are also more general arguments presented in \cite{Arfaei:1996zp} which ensure that the Polyakov-Wiegmann formula still holds.

Along the lines of our previous arguments we arrive again at a factorization of the partition function as

\begin{align}\begin{split}
	\mathcal Z = \mathcal Z^0 e^{-S(k)}.
\end{split}\end{align}
As $k$ is a classical field, the action in fact reduces to the simple form
\begin{align}\begin{split}
	S(k) &= \frac{\k}{2\pi i} \int \vev{k^\1 \P_\a k, k^\1 \P^\a k}\ .
\end{split}\end{align}
Using the explicit expression \cite{Nappi:1993ie}
\begin{align}\begin{split}
	k^\1\P_\a k &= \left(\cos u \P_\a a_k + \e_{jk}\sin u \P_\a a_j\right)P_k
				+ \P_\a u J + \left(\P_\a v + \2\e_{jk}\P_\a a_j a_k\right)T
\end{split}\end{align}
we arrive at
\begin{align}\begin{split}
	&\vev{k^\1 \P_\a k, k^\1 \P^\a k}
	= \P_\a a^i \P^\a a_i + b\P_\a u\P^\a u
		+ 2\P_\a u \P^\a v + \P_\a u \e_{jk}\P^\a a_j a_k
\end{split}\end{align}

After some partial integrations and use of the equations of motion, the action reduces to a boundary integral:
\begin{align}\begin{split}
\label{NWbdry}
	I&=I_0+I_1\\
	I_0&=\oint d\theta \Bigl\{
		a_1\P_n a_1 + a_2\P_n a_2 + v\P_n u + bu\P_n u \Bigr\}\\\
	I_1&=\oint d\theta \Bigl\{ \2 a_1a_2\P_\theta u + ua_1\P_\theta a_2
		\Bigr\}\ ,
\end{split}\end{align}
where $\P_n$ is the normal derivative at the boundary.

The flat space limit is achieved by the rescaling $(a_i, u, v)\to (\l a_i, \l u, \l v)$, 
$S\to \l^{-2} S$ and expansion around $\l=0$. The action takes the form 
$I=I^0 + \l I^1$ and this is exact, which means that no higher order corrections in $\lambda$ appear. 
The boundary theory can get contributions from the measure, which are higher order in $\lambda$. It can be checked though that the Jacobian is $J=1+{\cal O}(\lambda^2)$. Thus there are no first order contributions to the action coming from the measure.
The  shift $I_1$ in (\ref{NWbdry}) then implements the open string interaction that parametrizes the Nappi-Witten closed string background in background independent string field theory.

\section{Discussion}

Tachyon condensation processes have experienced much attention since they provide examples where closed strings are assumed to appear in open string theory. According to Sen's conjectures \cite{Sen:1999mg, Sen:2004nf} the condensation $T\to T_0$ will lead to a new vacuum of the theory, where on the way open string degrees of freedom are removed. In cases where all of them are removed, the vacuum must be a closed string vacuum. Any excitations of fields at the point $T_0$ must be closed string excitations, even when formulated originally in terms of open string fields.

A similar phenomenon is observed here. New couplings are added to the original theory, which correspond to non-local operators with respect to the original fields. But still, they can be expressed by the data contained in the open string map. By construction these couplings are connected to shifts in the closed string background field. Finally the observed condensation of a flat 3-brane to a spherical 2-brane gives further  support to the idea, that indeed closed strings in OSFT are described by non-local boundary operators. 
Thus we have given evidence that within 
the framework of BSFT different closed string $\sigma$-model backgrounds can be equivalently 
described in terms of non-local open string backgrounds. 

It should be stressed, however, that the factorization property of the $\sigma$-model partition function, which is the basic ingredient for the open-closed correspondence described here has not been proven to hold in general. While we have shown it to hold for group manifolds described by WZW models it does not appear to be valid in general for gauged WZW models \cite{DIPLOMA}. In particular, it does not hold in unmodified form for Liouville theory which is a gauged $SL(2,R)$ WZW model. There one finds that  
\be
\frac{d}{d\tau}\log[{\cal Z}^0_\tau]=\oint d\theta X_b(f)\<\partial_n X_0\>_\tau\,,\qquad{\cal Z}^0_\tau = \int D[X_0] e^{-I[X_0+\tau X_b]}\ .
\ee
The proper interpretation of the coupling of bulk - and boundary degrees of freedom  in Liouville theory is not completely understood at present. On the other hand there are reasons to believe that the factorization property will hold  for the class of CFTs which realize a chiral current algebra.

\subsection*{Acknowledgments }
 
We would like to thank Samson Shatashvili for collaboration. The 
work of I.S. was supported in parts by the Transregio TR-33 "Dark Energy" 
and the Excellence Cluster ÓOrigin and Structure of the UniverseÓ of the DFG.

\end{document}